\documentstyle[epsfig,12pt]{article}
\def \bea{\begin{eqnarray}}
\def \beq{\begin{equation}}

\def \br{${\cal B}$}
\def \bra#1{\langle #1 |}

\def \eea{\end{eqnarray}}
\def \eeq{\end{equation}}

\def \ket#1{| #1 \rangle}

\def \pr{\parallel}

\begin{document}
\Large
\centerline {\bf Production of the $\eta_b(nS)$ states 
\footnote{Enrico Fermi Institute preprint EFI 01-10, hep-ph/0104253.
To be published in Physical Review D.}}
\normalsize
\bigskip
 
\centerline{Stephen Godfrey~\footnote{godfrey@physics.carleton.ca}}
\centerline{\it Ottawa-Carleton Institute for Physics}
\centerline{\it Department of Physics, Carleton University}
\centerline{\it 1125 Colonel By Drive, Ottawa, ON K1S 5B6, Canada}
\smallskip
\centerline{and}
\smallskip
\centerline{Jonathan L. Rosner~\footnote{rosner@hep.uchicago.edu}}
\centerline {\it Enrico Fermi Institute and Department of Physics}
\centerline{\it University of Chicago, 5640 S. Ellis Avenue, Chicago, IL 60637}
\bigskip
 
\begin{quote}

The rates for magnetic dipole (M1) transitions $\Upsilon(nS) \to \eta_b(n'S) +
\gamma$, $n' \le n$, are compared.  The photon energies for allowed ($n' = n$)
M1 transitions are very small, so hindered ($n' < n$) transitions could be
more favorable for discovering the $\eta_b(1S,2S)$.  The question then
arises whether $\Upsilon(2S)$ or $\Upsilon(3S)$ is a better source of
$\eta_b(1S)$.  Whereas one nonrelativistic model favors $\eta_b(1S)$ production
from $\Upsilon(2S)$, this advantage is lost when relativistic corrections
are taken into account, and is not common to all sets of wave functions even
in the purely nonrelativistic limit.  Thus, the prospects for discovering
$\eta_b(1S)$ in $\Upsilon(3S)$ radiative decays could be
comparable to those in $\Upsilon(2S)$ decays.  We also discuss a
suggestion for discovering the $\eta_b$ via $\Upsilon(3S) \to h_b(^1P_1) \pi
\pi$, followed by $h_b \to \eta_b \gamma$.

\end{quote}
\bigskip

\noindent
PACS Categories:  14.40.Gx, 13.20.Gd, 13.40.Hq, 12.39.Ki

\bigskip

The $\Upsilon$ $b \bar b$ resonances have a rich spectroscopy \cite{revs}.
The spin-triplet S-wave levels $\Upsilon(nS)$ with $J^{PC} = 1^{--}$ are
produced by virtual photons in hadronic or $e^+ e^-$ interactions, and then
can undergo electric dipole (E1) transitions to the spin-triplet P-wave levels.
However, to reach the spin-singlet S-wave levels $\eta_b(nS)$
from the easily-produced $1^{--}$ states, it is necessary
to use either favored magnetic dipole (M1) transitions with very small photon
energy, or hindered M1 transitions with change of principal quantum number.
No spin-singlet $b \bar b$ levels have yet been seen.
The mass splitting between the singlet and triplet states is a key test
of the applicability of perturbative quantum chromodynamics (PQCD) to
the $b \bar b$ system \cite{NPT,FY} and is a useful check of lattice QCD 
results \cite{lattice}.

In this note we review some predictions for M1 transitions
\cite{JR83,MR,MB,ZB,GOS,GI,PTN,ZSG,EQ,LNR,UK,BSV} from the
$\Upsilon$ ($n^3S_1$) levels to the $\eta_b$ ($n'^1S_0$) states.  The photon
energies for allowed ($n'=n$) transitions are very small, so the hindered
($n'<n$) transitions could offer better prospects for discovering the
spin-singlet states.  The question then arises whether $\Upsilon(2S)$ or
$\Upsilon(3S)$ is a better source of $\eta_b(1S)$.  This question has taken on
renewed interest because of plans of the CLEO Collaboration at the Cornell
Electron Storage Ring (CESR) to increase their sample of data at the
$\Upsilon(3S)$ and possibly the $\Upsilon(2S)$ resonance.

The answer is very dependent on good knowledge of relativistic corrections
\cite{ZB,GOS,GI,ZSG,LNR,Sucher}.  In the absence of such corrections
the only source of a non-zero overlap between $nS$ and $1S$ wave functions
in the hindered transitions is the variation with $r$ of $j_0(kr/2)$, where $k$
is the photon energy in the rest frame of the decaying particle and $j_0(x)
\equiv (\sin x)/x$.  In some (e.g., \cite{JR83}) but not other (e.g, \cite{GI})
treatments, the matrix element of $j_0(kr/2)$ is much larger between $2S$ and
$1S$ than between $3S$ and $1S$ wave functions.  This hierarchy is largely
obliterated when relativitistic corrections are taken into account.  The most
important of these appears to be the difference between $^1S_0$ and $^3S_1$
wave functions due to the hyperfine interaction.  The attractive spin-spin
interaction in the $^1S_0$ states causes the wave function to be drawn
closer to the origin, leading to much less difference between the branching
ratios for $\Upsilon(2S) \to \gamma \eta_b$ and $\Upsilon(3S) \to \gamma
\eta_b$.  More recently one group has pointed out a crucial role for
exchange currents \cite{LNR}, which leads to very different conclusions from
those obtained previously.

The rates for magnetic dipole transitions in quarkonium ($Q \bar Q$) bound
states are given in the nonrelativistic approximation by \cite{Nov,JDJ}
\beq \label{eqn:rate}
\Gamma(^3S_1 \to ^1S_0 + \gamma) = \frac{4}{3} \alpha \frac{e_Q^2}{m_Q^2}
I^2 k^3~~,
\eeq
where $\alpha = 1/137.036$ is the fine-structure constant, $e_Q$ is the
quark charge in units of $|e|$ ($-1/3$ for $Q=b$), and $m_Q$ is the quark
mass (which we shall take equal to 4.8 GeV/$c^2$).  In all our discussions
we shall assume a normal magnetic moment of the $b$ quark.  The overlap
integral $I$ is defined by
\beq
I = \bra{f} j_0(kr/2) \ket{i}~~~.
\eeq

\begin{table}
\caption{Predictions for hyperfine splittings between $n^3S_1$ and $n^1S_0$
$b \bar b$ levels, and corresponding predicted branching ratios \br~for favored
M1 transitions. Overlap integrals have been set equal to unity except in the
second-to-last row.}
\begin{center}
\begin{tabular}{c c c c c c c} \hline
Reference & \multicolumn{2}{c}{$n=1$} & \multicolumn{2}{c}{$n=2$}
          & \multicolumn{2}{c}{$n=3$} \\ \hline
          & $\Delta M$ & \br & $\Delta M$ & \br & $\Delta M$ & \br \\
          & (MeV) & ($10^{-4}$) & (MeV) & ($10^{-4}$) & (MeV) & ($10^{-4}$) \\
\hline
MR83 \cite{JR83,MR} &  57 & 1.7 & 26 & 0.19 & 19 & 0.12 \\
MB83 \cite{MB}      & 100 & 8.9 & 40 & 0.68 & 31 & 0.53 \\
GOS84 \cite{GOS} (a) & 67 & 2.7 & 31 & 0.32 & $-3$ & 0 \\
GOS84 \cite{GOS} (b) & 78 & 4.2 & 37 & 0.54 & 27 & 0.35 \\
GI85 \cite{GI} (c)  &  63 & 2.2 & 27 & 0.21 & 18 & 0.10 \\
PTN86 \cite{PTN}    &  35 & 0.38 & 19 & 0.07 & 15 & 0.06 \\
FY99 \cite{FY}	    & 53 & 1.3 & (d) & (d) & (d) & (d) \\
ZSG91 \cite{ZSG}    &  48 & 0.99 & 28 & 0.23 & (d) & (d) \\
EQ94 \cite{EQ}      &  87 & 5.9 & 44 & 0.91 & 41 & 1.2 \\
LNR99 \cite{LNR}    &  79 & 4.4 & 44 & 0.91 & 35 & 0.76 \\
LNR99 \cite{LNR} (e) &  79 & 3.6 & 44 & 0.63 & 35 & 0.50 \\
UKQCD00 \cite{UK}   &  42 & 0.66 & (d) & (d) & (d) & (d) \\
BSV01 \cite{BSV}    & 36--55 & 0.4--1.5 & (d) & (d) & (d) & (d) \\ \hline
\end{tabular}
\end{center}
\leftline{\qquad \qquad $^a$ Scalar confining potential (favored by fit
to P-waves).}
\leftline{\qquad \qquad $^b$ Vector confining potential.}
\leftline{\qquad \qquad $^c$ The splittings are based on masses 
rounded to 1 MeV, not the results}
\leftline{\qquad \qquad rounded to 10 MeV as given in Ref.\ \cite{GI}.}
\leftline{\qquad \qquad $^d$ Not quoted. $^e$ Results for fully relativistic
calculation.}
\end{table}
\bigskip

We summarize in Table I some predictions for
mass splittings between $n^3S_1$ and $n^1S_0$ $b \bar b$ levels and the
corresponding branching ratios for $\Upsilon(nS) \to \eta_b(nS) + \gamma$
entailed by Eq.~(\ref{eqn:rate}) assuming unit overlap integral, $I=1$.
We take the total widths of the $\Upsilon(nS)$ levels to be
\cite{PDG} $\Gamma_{\rm tot}[\Upsilon(1S,2S,3S)] = (52.5,44,26.3)$ keV.
For the low-energy favored M1 transitions, the photon energies are nearly
the same as the mass splittings.  The wide variation in predicted hyperfine
splittings leads to considerable uncertainty in predicted rates
for these transitions.

\begin{table}
\caption{Predictions for overlap integrals and branching ratios in hindered M1
transitions between $n^3S_1$ and $n'^1S_0$ $b \bar b$ levels, {\it neglecting}
relativistic corrections.}
\begin{center}
\begin{tabular}{c c c c c c c} \hline
          & \multicolumn{2}{c}{$n=2,~n'=1$} & \multicolumn{2}{c}{$n=3,~n'=1$} &
\multicolumn{2}{c}{$n=3,~n'=2$} \\
          & \multicolumn{2}{c}{$k = 600$ MeV} & \multicolumn{2}{c}{$k = 908$
MeV} & \multicolumn{2}{c}{$k = 352$ MeV} \\ \hline
Reference & $|I|$ & \br & $|I|$ & \br & $|I|$ & \br \\
          & & ($10^{-4}$) & & ($10^{-4}$) & & ($10^{-4}$) \\ \hline
JR83 \cite{JR83} & 0.02 & 0.92 & $< 0.002$ & $< 0.06$ & $\sim 0.005$ &
0.01--0.04 \\
GI85 \cite{GI00} & 0.017 & 0.67 & 0.0070 & 0.65 & 0.018 & 0.25 \\
ZSG91 \cite{ZSG} (a) & 0.069 & 11.0 & (b) & (b) & (b) & (b) \\
ZSG91 \cite{ZSG} (c) & 0.022 & 1.11 & (b) & (b) & (b) & (b) \\
LNR99 \cite{LNR} &  (b)  & 0.78 &   (b)  & 1.1 &  (b) & 0.54 \\ \hline 
\end{tabular}
\end{center}
\leftline{\qquad \qquad $^a$ Scalar-vector exchange potential of Ref.\
\cite{GRS}. $^b$ Not quoted.}
\leftline{\qquad \qquad $^c$ Scalar exchange potential of Ref.\ \cite{GRS}.}
\end{table}

For the higher-energy hindered M1 transitions, the expected photon energies
$k = (M_i^2 - M_f^2)/(2 M_i)$ are not so sensitive to hyperfine splittings. 
On the basis of present experimental values for the $\Upsilon(1S,2S,3S)$
masses \cite{PDG} of (9460,10023,10355) MeV/$c^2$ and the hyperfine splittings
predicted in Ref.~\cite{JR83}, the masses for the $\eta_b$ levels from are
predicted to be $M(1S,2S,3S) = (9403,9997,10336)$ MeV/$c^2$, a representative
set which we shall take in further calculations.  We then compare two
predictions for overlap integrals and branching ratios in Table II, {\it
taking into account only the expectation value of the spherical Bessel
function $j_0(kr/2)$ between initial and final states}.  While the overlaps
for the $n=2,~n'=1$ transition are similar, the other overlaps differ
significantly, suggesting that they may be sensitive to small details of
the potential (as in the case of two different potentials \cite{GRS} utilized
by Ref.\ \cite{ZSG}).

\begin{table}
\caption{Predictions for overlap integrals and branching ratios in hindered M1
transitions between $n^3S_1$ and $n'^1S_0$ $b \bar b$ levels, taking into
account relativistic corrections.}
\begin{center}
\begin{tabular}{c c c c c c c} \hline
          & \multicolumn{2}{c}{$n=2,~n'=1$} & \multicolumn{2}{c}{$n=3,~n'=1$} &
\multicolumn{2}{c}{$n=3,~n'=2$} \\
          & \multicolumn{2}{c}{$k = 600$ MeV} & \multicolumn{2}{c}{$k = 908$
MeV} & \multicolumn{2}{c}{$k = 352$ MeV} \\ \hline
Reference & $|I|$ & \br & $|I|$ & \br & $|I|$ & \br \\
          & & ($10^{-4}$) & & ($10^{-4}$) & & ($10^{-4}$) \\ \hline
ZB83 \cite{ZB} & 0.080 & 15 & 0.041 & 22 & 0.095 & 7.0 \\
GOS84 \cite{GOS} (a) & (b) & 7.9 & (b) & (b) & (b) & (b) \\
GOS84 \cite{GOS} (c) & (b) & 5.4 & (b) & (b) & (b) & (b) \\
GI85 \cite{GI} (d) & 0.057 & 7.4 & 0.029 & 11 & 0.054 & 2.2 \\
GI85 \cite{GI00} (e) & 0.081 & 13 & 0.043 & 25 & 0.078 & 4.7 \\
ZSG91 \cite{ZSG} (f) & 0.025 & 1.4 & (b) & (b) & (b) & (b) \\
ZSG91 \cite{ZSG} (g) & 0.001 & $\sim 0$ & (b) & (b) & (b) & (b) \\
LNR99 \cite{LNR} (h) & (b) & 0.46 & (b) & 1.4 & (b) & 0.13 \\
LNR99 \cite{LNR} (i) & (b) & 0.05 & (b) & 0.05 & (b) & 0.40 \\ \hline
\end{tabular}
\end{center}
\leftline{\qquad \qquad $^a$ Scalar confining potential.  $^b$ Not quoted.}
\leftline{\qquad \qquad $^c$ Vector confining potential.}
\leftline{\qquad \qquad $^d$ Based on quoted transition moments.}
\leftline{\qquad \qquad $^e$ Based on matrix elements between $^3S_1$
and $^1S_0$ wave functions.}
\leftline{\qquad \qquad $^f$ Scalar-vector confining potential of Ref.\
\cite{GRS}.}
\leftline{\qquad \qquad $^g$ Scalar confining potential of Ref.\
\cite{GRS}.}
\leftline{\qquad \qquad $^h$ Without exchange current.
$^i$ With exchange current.}
\end{table}

When relativistic corrections are taken into account, the results are
as shown in Table III.  Several calculations \cite{ZB,GI,GI00} predict large
branching ratios for all sets of hindered transitions.  
There appears to be no particular advantage
in searching for the $\eta_b(1S)$ state at the $\Upsilon(2S)$; the branching
ratio from the $\Upsilon(3S)$ is predicted to be slightly larger, partly
compensating for the lower production cross section of the $\Upsilon(3S)$.
(The cross sections for $e^+ e^- \to \Upsilon(2S)$ and $e^+ e^- \to
\Upsilon(3S)$ are measured to be about 6.6 and 4 nb, respectively
\cite{CL2S,CL3S,CU3S}, for final states other than lepton pairs.)

The calculation of Ref.\ \cite{GI00} is based entirely on the distortion of
the spin-singlet wave functions by the strong hyperfine attraction.  
The spin-independent
potential consists of a short distance Coulomb-type Lorentz vector 
and a long distance linear Lorentz scalar interaction. The hyperfine 
interaction is included directly in the Hamiltonian by smearing the 
relative coordinate over distances of the order of the inverse quark 
masses which has the consequence of taming the singularities present 
in the Breit-Fermi interaction.  The resulting 
wave functions are compared with the corresponding spin-triplet wave functions
in Fig.\ 1.  The stronger peaking near the origin of the singlet wave
functions is clearly visible.

The best of present
data \cite{CL2S,CL3S,CU3S} may not be adequate to confront such predictions,
and no published limits are reported at present.  For example, at the
$\Upsilon(2S)$, the transition to the $\chi_{b0}(1P)$ and a 162 MeV photon
corresponds to a signal of $8637 \pm 1274$ events in CLEO data \cite{CL2S},
for a branching ratio of $(3.4 \pm 0.5 \pm 0.6)\%$.  Although a considerable
extrapolation is needed to anticipate the signal of a 600 MeV photon (since
the spectrum for that energy is not published), it is likely that such a
photon emitted
with a branching ratio of $10^{-3}$ would be lost in the combinatorial
background associated with neutral pions.  Similarly, in cascade decays
from the $\Upsilon(3S)$ to the $\chi'_{bJ}(2P)$ states followed by $\chi'_{bJ}
(2P) \to \Upsilon(1S) \gamma$, the 770 MeV photon corresponds to a signal
of $1994 \pm 150$ events in CUSB data \cite{CU3S}, for a branching ratio
of $(2.0 \pm 0.2 \pm 0.2)\%$.  This is to be compared with the sought-for
signal of a $\sim 900$ MeV photon emitted with a branching ratio of about
$2 \times 10^{-3}$.  The corresponding spectrum for the CLEO $\Upsilon(3S)$
data, based on a smaller sample, shows similar features \cite{TS}.

\begin{figure}
\centerline{\epsfysize = 5in \epsffile{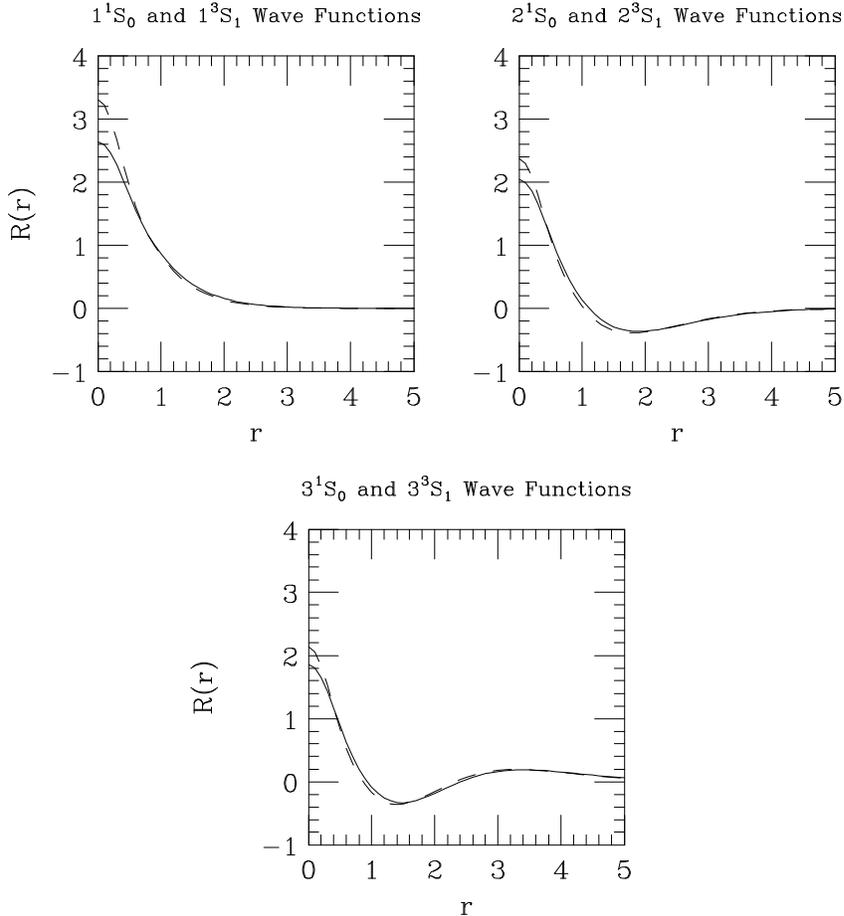}}
\caption{Comparison of spin-singlet (dashed) and spin-triplet (solid)
wave functions for $nS$ $b \bar b$ levels.}
\end{figure}

We conclude with some remarks on the production of the $\eta_b(1S)$ level
through the decay $\Upsilon(3S) \to h_b(1^1P_1) \pi \pi$, which is predicted
by Kuang and Yan \cite{KY} to have a branching ratio of about 0.1--1\%.
[Voloshin \cite{Vol} finds a much smaller value for this quantity, less
than $10^{-4}$, and suggests observing instead the isospin-violating
transition $\Upsilon(3S) \to h_b(1^1P_1) \pi^0$, for which he predicts
a branching ratio of $10^{-3}$.]  The subsequent electric dipole decay $h_b \to
\eta_b(1S) + \gamma$ is predicted \cite{KY}
to have a branching ratio approaching 50\%.  The mass of $h_b$ is expected to
be not far from the center-of-gravity of the $1^3P_J$ $\chi_b$ levels,
or about 9.90 GeV/$c^2$, so the photon should have an energy of about 485
MeV in the $h_b$ rest frame.  The CUSB Collaboration has searched for the
$h_b(1^1P_1) \pi \pi$ signature and is able to
place an upper limit at 90\% c.l. \cite{CU3S} of $< 0.45\%$ on the combined
branching ratio for an $\Upsilon(1S)$--$\eta_b(1S)$ splitting ranging between
50 and 110 MeV.  The CLEO Collaboration \cite{CL3Sh} places an inclusive
upper limit of ${\cal B}[\Upsilon(3S) \to \pi^+ \pi^- h_b] < 0.18\%$ at
90\% c.l. for $M_{h_b} = 9.900$ GeV/$c^2$ and 90\% c.l. upper limits in the
range of 0.1\% for the cascade ${\cal B}[\Upsilon(3S) \to \pi^+ \pi^- h_b
\to \pi^+ \pi^- \eta_b \gamma]$, with $M_{h_b} = 9.900 \pm 0.003$ GeV/$c^2$
and a photon energy between 434 and 466 MeV.

It appears that the richness of transitions to $\eta_b(nS)$ levels available
in decays of the $\Upsilon(3S)$ make it a promising initial candidate for
enhanced searches for the elusive spin-singlet levels.

[Note added:  We thank R. Faustov and V. Galkin for calling attention to
their work on the relativistic quark model, e.g., Refs. \cite{FG,EFG}.  They
predict $\eta_b$ and $\eta'_b$ to lie 60 and 30 MeV below their respective
$^3S_1$ partners, and ${\cal B}(\Upsilon(1S) \to \eta_b \gamma) = 0.88
\times 10^{-4}$, ${\cal B}(\Upsilon(2S) \to \eta_b \gamma) = 1.6 \times
10^{-4}$ (private communication).]

We thank R. S. Galik for asking the question which led to this investigation,
and for extensive discussions.  We also thank T. Skwarnicki for helpful
advice, and S. F. Tuan for reminding us of the distinction between Refs.\
\cite{KY} and \cite{Vol}.  This work was supported in part by the United
States Department of Energy through Grant No.\ DE FG02 90ER40560
and the Natural Sciences and Engineering Research Council of Canada.

\def \ajp#1#2#3{Am.\ J. Phys.\ {\bf#1}, #2 (#3)}
\def \apny#1#2#3{Ann.\ Phys.\ (N.Y.) {\bf#1}, #2 (#3)}
\def \app#1#2#3{Acta Phys.\ Polonica {\bf#1}, #2 (#3)}
\def \arnps#1#2#3{Ann.\ Rev.\ Nucl.\ Part.\ Sci.\ {\bf#1}, #2 (#3)}
\def \art{and references therein}
\def \cmts#1#2#3{Comments on Nucl.\ Part.\ Phys.\ {\bf#1}, #2 (#3)}
\def \cn{Collaboration}
\def \cp89{{\it CP Violation,} edited by C. Jarlskog (World Scientific,
Singapore, 1989)}
\def \efi{Enrico Fermi Institute Report No.\ }
\def \epjc#1#2#3{Eur.\ Phys.\ J. C {\bf#1}, #2 (#3)}
\def \f79{{\it Proceedings of the 1979 International Symposium on Lepton and
Photon Interactions at High Energies,} Fermilab, August 23-29, 1979, ed. by
T. B. W. Kirk and H. D. I. Abarbanel (Fermi National Accelerator Laboratory,
Batavia, IL, 1979}
\def \hb87{{\it Proceeding of the 1987 International Symposium on Lepton and
Photon Interactions at High Energies,} Hamburg, 1987, ed. by W. Bartel
and R. R\"uckl (Nucl.\ Phys.\ B, Proc.\ Suppl., vol.\ 3) (North-Holland,
Amsterdam, 1988)}
\def \ib{{\it ibid.}~}
\def \ibj#1#2#3{~{\bf#1}, #2 (#3)}
\def \ichep72{{\it Proceedings of the XVI International Conference on High
Energy Physics}, Chicago and Batavia, Illinois, Sept. 6 -- 13, 1972,
edited by J. D. Jackson, A. Roberts, and R. Donaldson (Fermilab, Batavia,
IL, 1972)}
\def \ijmpa#1#2#3{Int.\ J.\ Mod.\ Phys.\ A {\bf#1}, #2 (#3)}
\def \ite{{\it et al.}}
\def \jhep#1#2#3{JHEP {\bf#1}, #2 (#3)}
\def \jpb#1#2#3{J.\ Phys.\ B {\bf#1}, #2 (#3)}
\def \lg{{\it Proceedings of the XIXth International Symposium on
Lepton and Photon Interactions,} Stanford, California, August 9--14 1999,
edited by J. Jaros and M. Peskin (World Scientific, Singapore, 2000)}
\def \lkl87{{\it Selected Topics in Electroweak Interactions} (Proceedings of
the Second Lake Louise Institute on New Frontiers in Particle Physics, 15 --
21 February, 1987), edited by J. M. Cameron \ite~(World Scientific, Singapore,
1987)}
\def \kdvs#1#2#3{{Kong.\ Danske Vid.\ Selsk., Matt-fys.\ Medd.} {\bf #1},
No.\ #2 (#3)}
\def \ky85{{\it Proceedings of the International Symposium on Lepton and
Photon Interactions at High Energy,} Kyoto, Aug.~19-24, 1985, edited by M.
Konuma and K. Takahashi (Kyoto Univ., Kyoto, 1985)}
\def \mpla#1#2#3{Mod.\ Phys.\ Lett.\ A {\bf#1}, #2 (#3)}
\def \nat#1#2#3{Nature {\bf#1}, #2 (#3)}
\def \nc#1#2#3{Nuovo Cim.\ {\bf#1}, #2 (#3)}
\def \nima#1#2#3{Nucl.\ Instr.\ Meth. A {\bf#1}, #2 (#3)}
\def \np#1#2#3{Nucl.\ Phys.\ {\bf#1}, #2 (#3)}
\def \npbps#1#2#3{Nucl.\ Phys.\ B Proc.\ Suppl.\ {\bf#1}, #2 (#3)}
\def \os{XXX International Conference on High Energy Physics, Osaka, Japan,
July 27 -- August 2, 2000}
\def \PDG{Particle Data Group, D. E. Groom \ite, \epjc{15}{1}{2000}}
\def \pisma#1#2#3#4{Pis'ma Zh.\ Eksp.\ Teor.\ Fiz.\ {\bf#1}, #2 (#3) [JETP
Lett.\ {\bf#1}, #4 (#3)]}
\def \pl#1#2#3{Phys.\ Lett.\ {\bf#1}, #2 (#3)}
\def \pla#1#2#3{Phys.\ Lett.\ A {\bf#1}, #2 (#3)}
\def \plb#1#2#3{Phys.\ Lett.\ B {\bf#1}, #2 (#3)}
\def \pr#1#2#3{Phys.\ Rev.\ {\bf#1}, #2 (#3)}
\def \prc#1#2#3{Phys.\ Rev.\ C {\bf#1}, #2 (#3)}
\def \prd#1#2#3{Phys.\ Rev.\ D {\bf#1}, #2 (#3)}
\def \prl#1#2#3{Phys.\ Rev.\ Lett.\ {\bf#1}, #2 (#3)}
\def \prp#1#2#3{Phys.\ Rep.\ {\bf#1}, #2 (#3)}
\def \ptp#1#2#3{Prog.\ Theor.\ Phys.\ {\bf#1}, #2 (#3)}
\def \rmp#1#2#3{Rev.\ Mod.\ Phys.\ {\bf#1}, #2 (#3)}
\def \rp#1{~~~~~\ldots\ldots{\rm rp~}{#1}~~~~~}
\def \rpp#1#2#3{Rep.\ Prog.\ Phys.\ {\bf#1}, #2 (#3)}
\def \sing{{\it Proceedings of the 25th International Conference on High Energy
Physics, Singapore, Aug. 2--8, 1990}, edited by. K. K. Phua and Y. Yamaguchi
(Southeast Asia Physics Association, 1991)}
\def \slc87{{\it Proceedings of the Salt Lake City Meeting} (Division of
Particles and Fields, American Physical Society, Salt Lake City, Utah, 1987),
ed. by C. DeTar and J. S. Ball (World Scientific, Singapore, 1987)}
\def \slac89{{\it Proceedings of the XIVth International Symposium on
Lepton and Photon Interactions,} Stanford, California, 1989, edited by M.
Riordan (World Scientific, Singapore, 1990)}
\def \smass82{{\it Proceedings of the 1982 DPF Summer Study on Elementary
Particle Physics and Future Facilities}, Snowmass, Colorado, edited by R.
Donaldson, R. Gustafson, and F. Paige (World Scientific, Singapore, 1982)}
\def \smass90{{\it Research Directions for the Decade} (Proceedings of the
1990 Summer Study on High Energy Physics, June 25--July 13, Snowmass, Colorado),
edited by E. L. Berger (World Scientific, Singapore, 1992)}
\def \tasi{{\it Testing the Standard Model} (Proceedings of the 1990
Theoretical Advanced Study Institute in Elementary Particle Physics, Boulder,
Colorado, 3--27 June, 1990), edited by M. Cveti\v{c} and P. Langacker
(World Scientific, Singapore, 1991)}
\def \yaf#1#2#3#4{Yad.\ Fiz.\ {\bf#1}, #2 (#3) [Sov.\ J.\ Nucl.\ Phys.\
{\bf #1}, #4 (#3)]}
\def \zhetf#1#2#3#4#5#6{Zh.\ Eksp.\ Teor.\ Fiz.\ {\bf #1}, #2 (#3) [Sov.\
Phys.\ - JETP {\bf #4}, #5 (#6)]}
\def \zpc#1#2#3{Zeit.\ Phys.\ C {\bf#1}, #2 (#3)}
\def \zpd#1#2#3{Zeit.\ Phys.\ D {\bf#1}, #2 (#3)}

\end{document}